\documentclass[12pt,showpacs,aps,nofootinbib]{revtex4}
\usepackage{amsmath,amssymb,graphicx,epsfig,color,ulem,cancel,xcolor,datetime}
\def\comment#1{}

\def\nn{\nonumber}
\def\rnd{\partial}
\def\ds{\mathrm{dS}}
\def\x{\mathbf{x}}

\def\vin{\mathrm{in}}
\def\vout{\mathrm{out}}
\def\sem{\mathrm{sem}}
\def\reg{\mathrm{reg}}
\def\ky{\mathbf{k}_{y\!\!\!\!\diagup}}
\def\dky{k_{y\!\!\!\!\diagup}}
\def\py{\mathbf{p}_{y\!\!\!\!\diagup}}
\def\dst{\text{dS}_{2}}
\def\dsf{\text{dS}_{4}}
\def\A{\mathrm{A}}
\begin{document}
%%%%%%%%%%%%%%%%%%%%%%%%%%%%%%%%%%%%%%%%%%%%%%%%%%%%%%%%%%%%%%%%%%%%%%%%%%%%%%%%%%%%%%%%%%%%%%%%%%%%%%%%%%%%%%%%%%%%%%%%%%%%%%%%%%%%%%%%%%%%%%%%%%%%%%%%%%%%%%%
%%%%%%%%%%%%%%%%%%%%%%%%%%%%%%%%%%%%%%%%%%%%%%%%%%%%%%%%%%%%%%%%%%%%%%%%%%%%%%%%%%%%%%%%%%%%%%%%%%%%%%%%%%%%%%%%%%%%%%%%%%%%%%%%%%%%%%%%%%%%%%%%%%%%%%%%%%%%%%%
\title{Effect of a magnetic field on Schwinger mechanism in de~Sitter spacetime}
\author{Ehsan~Bavarsad$^{1}$}\email{bavarsad@kashanu.ac.ir}
\author{Sang~Pyo~Kim$^{2,3}$}
\author{Cl\'{e}ment~Stahl$^{4}$}
\author{She-Sheng~Xue$^{5,6}$}
\affiliation{$^{1}$Department of Physics, University of Kashan, 8731753153, Kashan, Iran \\
$^{2}$Department of Physics, Kunsan National University, Kunsan 54150, Korea \\
$^{3}$Center for Relativistic Laser Science, Institute for Basic Science, Gwangju 61005, Korea \\
$^{4}$Instituto de F\'{\i}sica, Pontificia Universidad Cat\'{o}lica de Valpara\'{\i}so, Casilla 4950, Valpara\'{\i}so, Chile \\
$^{5}$ICRANet, Piazzale della Repubblica 10, 65122 Pescara, Italy \\
$^{6}$Dipartimento di Fisica, Universita di Roma ``La Sapienza'', Piazzale Aldo Moro 5, 00185 Rome, Italy  }

%\date{\blue{Revised Version:~\today~at~\currenttime}}

\begin{abstract}
We investigate the effect of a constant magnetic field background on the scalar QED pair production in a four-dimensional de~Sitter spacetime ($\dsf$).
We have obtained the pair production rate which agrees with the known Schwinger result in the limit of Minkowski spacetime and with the Hawking radiation in de~Sitter spacetime (dS) in the zero electric field limit.
Our results describe how the cosmic magnetic field affects the pair production rate in cosmological setups.
In addition, using the zeta function regularization scheme we have calculated the induced current and examined the effect of a magnetic field on the vacuum expectation value of the current operator.
We find that, in the case of a strong electromagnetic background the current responds as $E\cdot B$, while in the infrared regime, it responds as $B/E$, which leads to a phenomenon of infrared hyperconductivity.
These results of the induced current have important applications for the cosmic magnetic field evolution.
\end{abstract}

\maketitle
%\tableofcontents
%%%%%%%%%%%%%%%%%%%%%%%%%%%%%%%%%%%%%%%%%%%%%%%%%%%%%%%%%%%%%%%%%%%%%%%%%%%%%%%%%%%%%%%%%%%%%%%%%%%%%%%%%%%%%%%%%%%%%%%%%%%%%%%%%%%%%%%%%%%%%%%%%%%%%%%%%%%%%%%
%%%%%%%%%%%%%%%%%%%%%%%%%%%%%%%%%%%%%%%%%%%%%%%%%%%%%%%%%%%%%%%%%%%%%%%%%%%%%%%%%%%%%%%%%%%%%%%%%%%%%%%%%%%%%%%%%%%%%%%%%%%%%%%%%%%%%%%%%%%%%%%%%%%%%%%%%%%%%%%
%\iffalse
%\fi
%%%%%%%%%%%%%%%%%%%%%%%%%%%%%%%%%%%%%%%%%%%%%%%%%%%%%%%%%%%%%%%%%%%%%%%%%%%%%%%%%%%%%%%%%%%%%%%%%%%%%%%%%%%%%%%%%%%%%%%%%%%%%%%%%%%%%%%%%%%%%%%%%%%%%%%%%%%%%%%
%%%%%%%%%%%%%%%%%%%%%%%%%%%%%%%%%%%%%%%%%%%%%%%%%%%%%%%%%%%%%%%%%%%%%%%%%%%%%%%%%%%%%%%%%%%%%%%%%%%%%%%%%%%%%%%%%%%%%%%%%%%%%%%%%%%%%%%%%%%%%%%%%%%%%%%%%%%%%%%
\section{\label{sec:intro}Introduction}
A fascinating effect in quantum field theory is the Schwinger effect \cite{Gelis:2015kya}: the creation of pairs out of the vacuum in the presence
of a background electromagnetic field. While it was Sauter \cite{Sauter:1931zz}, Heisenberg, and student Euler \cite{Heisenberg:1935qt} who investigated first this effect, the history has remembered Schwinger who revisited their works some 20 years later \cite{Schwinger:1951nm}.
Despite being a very useful tool for the theoretical understanding of quantum field theory and for the development of powerful calculation techniques in
strong field background, the Schwinger effect has so far not been detected in laboratories experiments.
The production of electron-positron pairs, however, was realized in an experiment, in which high energy gammas scatter with a Coulomb potential
\cite{Burke:1997ew}.
The main reason is that the Schwinger effect is exponentially suppressed unless the electric field reaches close enough a threshold electric field $E_{\mathrm{threshold}}\simeq 1.3\times 10^{18}$ V/m \cite{DiPiazza:2011tq}.
Aiming at detecting this effect a new idea is developing in the past years: changing the system under study and considering Schwinger effect in astrophysical and cosmological contexts where huge background fields could naturally be present \cite{Ruffini:2009hg}.
We will investigate in this paper the Schwinger effect in the $\dsf$ under the influence of both a constant electric field and a magnetic field
background.
\par

The Schwinger effect in dS has recently become an active field of research.
The seminal papers studied this effect in the two-dimensional de~Sitter spacetime ($\dst$) \cite{Frob:2014zka} and in $\dsf$ \cite{Kobayashi:2014zza}.
The one-loop vacuum polarization and Schwinger effect in a two-dimensional (anti-)de~Sitter spacetime was explicitly found and a thermal interpretation
was proposed for the Schwinger effect in Ref.~\cite{Cai:2014qba}.
The initial motivation of \cite{Frob:2014zka} was to use this framework to investigate bubble nucleation in the context of the multiverse proposal.
However, this toy model for pair creation turns out to have a wide range of applications, from constraining magnetogenesis scenarios
\cite{Kobayashi:2014zza}, investigating the ER=EPR conjecture via holographic setups \cite{Fischler:2014ama} to pair creation around charged black holes \cite{Chen:2012zn,Kim:2016dmm,Chen:2016caa} and baryogenesis \cite{Stahl:2016qjs}.
\par

These physical motivations lead to a series of papers in which the Schwinger mechanism has been investigated for various types of particles and spacetime dimensions. It was investigated whether the known equivalence between bosonic and fermionic particles with respect to the Schwinger effect holds in $\dst$ \cite{Stahl:2015gaa}.
Particles differentiate themselves only if one goes beyond the semiclassical limit and computes the current which, in turn, is a more physically relevant quantity to describe the Schwinger mechanism in curved spacetimes.
These results were generalized to $\dsf$ in \cite{Hayashinaka:2016qqn}. For bosons in $\dsf$, the results of \cite{Kobayashi:2014zza} were reinforced by \cite{Hayashinaka:2016dnt} which considered an alternative renormalization scheme and found the same results. In \cite{Geng:2017zad}, an alternative
method was employed: the uniform asymptotic method was used to derive new results for the Schwinger effect in $\dsf$; see also \cite{Sharma:2017ivh}.
In \cite{Bavarsad:2016cxh}, the Schwinger mechanism in three dimensions was explored as an example of odd dimension field theory in dS.
In all these works the gravitational field and electric field were assumed to be background fields whose variations due to backreactions are negligible during the typical time scale of pair creation. This approximation can be shown to hold for some range of the parameters.
However, taking a constant background field can only be seen as a toy model to understand some physical implications of pair creation, and in realistic models of inflation requiring quasi-dS, the backreaction effects both on the dS metric and on the background electric field should be taken into account.
In \cite{Bavarsad:2016cxh} and \cite{Stahl:2016geq}, it was shown that both the gravitational and electromagnetic field would be suppressed by the
Schwinger effect. In Refs.~\cite{Xue:2014kna,Xue:2015tmw}, it was pointed out that the quantum-gravity originated cosmological constant term
$\Lambda g^{\mu\nu}$ results in the creation of particle-antiparticle pairs and their fields whose energy-momentum tensor $T^{\mu\nu}_{\mathrm{M}}$ in
turn backreacts on $\Lambda g^{\mu\nu}$, and that these are important to understand the inflationary process in the early Universe and the dark-energy-matter interaction for $\Omega_{\Lambda}\sim\Omega_{\mathrm{M}}\sim\mathcal{O}(1)$ in the present Universe.
Recently it was argued that the dS was unstable due to quantum effects \cite{Markkanen:2016aes,Markkanen:2016jhg,Markkanen:2017abw}.
The idea is that a non-trivial Bogoliubov transformation leads, after decoherence, to a breaking of the dS invariance and therefore to a decrease of the cosmological constant.
\par

In this article, we propose to take one step further and add a constant magnetic field to a dS and an already present electric background.
This is a common generalization of a flat spacetime in which the analytic results have been known for long \cite{Schwinger:1951nm}, but the Schwinger
effect has never been properly investigated in dS. One motivation to consider a constant magnetic field in dS is the recent result that a constant
magnetic field is a stable configuration of dS in modified gravity theories \cite{Mukohyama:2016npi}.
The effect of a constant magnetic field exhibits diversity of the Schwinger mechanism compared to a pure electric field in dS.
And another possible reason of considering an electromagnetic field in the early Universe would come from the observation of blazars leading to a lower bound for the magnetic field in the intergalactic medium: $B>6\times 10^{-18}$ G \cite{Vovk:2011aa}.
The origin of these magnetic fields is now an open question in cosmology but two main scenarios are emerging: their origin is either after recombination
or primordial; see reviews \cite{Widrow:2002ud,Kandus:2010nw,Durrer:2013pga,Subramanian:2015lua}.
In the case of a primordial origin, just as for a scalar field, the vacuum fluctuations of the gauge field are amplified to larger scales. Once inflation comes to an end, the Universe becomes conductive, leading the electric field to vanish and the magnetic field to reside and evolve until the present epoch by the flux conservation. If the primordial origin of the currently observed magnetic field is adopted, it is necessary for inflation model builders to investigate physical effects due to the presence of an electromagnetic field, i.e, the Schwinger effect, which is the main topic of this paper.
\par

The effect of a magnetic field background on the scalar pair creation probability \cite{Moradi:2009zz} and the number density \cite{Haouat:2012dr} in the spatially flat Friedmann-Lemaitre-Robertson-Walker (FLRW) universes have been investigated.
In \cite{Moradi:2009zz}, the author showed that in the presence of a pure magnetic field background, i.e., in the absence of an electric field
background, the gravitational pair creation does not change in dS, whereas in a radiation dominated universe, a pure magnetic field background minimizes
the gravitational pair creation \cite{Haouat:2012dr}.
In holographic setups, the inclusion of a magnetic field on the Schwinger effect was investigated in \cite{Ghodrati:2015rta}.
It is, however, difficult to compare that result directly with the case of dS under consideration in this paper.
Adopting the perturbative QED approach in dS, the first order amplitude for the fermion production in a magnetic field has been analyzed in \cite{Crucean:2016eiq}; see also \cite{Nicolaevici:2016uzy,Crucean:2017pfg}.
The authors found that the fermion production is significant only at large expansion condition.
This paper aims at investigating the magnetic field influence on the Schwinger pair creation of charged scalars in $\dsf$, specifically, by computing the semiclassical decay rate and analysing the quantum vacuum expectation value of the current operator, which is equivalent to the exact one-loop approach including all one-loop diagrams.
\par

The organization of this paper is as follows. In Sec.~\ref{sec:KG}, we recall the main equations for charged scalars in a magnetic field as well as an electric field for the pair creation setup.
In Sec.~\ref{sec:Schwinger}, we compute the pair creation rate using a semiclassical approach to the exact one-loop.
In Sec.~\ref{sec:current}, we present an expression for the induced current and discuss several relevant limiting cases of different field intensities.
We draw some conclusions and future lines of research in Sec.~\ref{sec:concl}.
Appendix~\ref{app:zeta} contains some mathematical aspects of this work: some useful properties of the Riemann and Hurwitz zeta functions.
Eventually, in Appendix~\ref{app:reg}, the computation and regularization of the current have been reviewed.
%%%%%%%%%%%%%%%%%%%%%%%%%%%%%%%%%%%%%%%%%%%%%%%%%%%%%%%%%%%%%%%%%%%%%%%%%%%%%%%%%%%%%%%%%%%%%%%%%%%%%%%%%%%%%%%%%%%%%%%%%%%%%%%%%%%%%%%%%%%%%%%%%%%%%%%%%%%%%%%
%%%%%%%%%%%%%%%%%%%%%%%%%%%%%%%%%%%%%%%%%%%%%%%%%%%%%%%%%%%%%%%%%%%%%%%%%%%%%%%%%%%%%%%%%%%%%%%%%%%%%%%%%%%%%%%%%%%%%%%%%%%%%%%%%%%%%%%%%%%%%%%%%%%%%%%%%%%%%%%
\section{\label{sec:KG}Klein-Gordon Equation}
To study the Schwinger effect in $\dsf$, we consider the action of a complex scalar field coupled to a $U(1)$ gauge field as
\begin{equation}
S=\int d^{4}x\sqrt{-g}\Bigl[g^{\mu\nu}\bigl(\rnd_{\nu}-ie A_{\nu}\bigr)\varphi^{\ast}\bigl(\rnd_{\mu}+ie A_{\mu}\bigr)\varphi
-\bigl(m^{2}+\xi R\bigr)\varphi\varphi^{\ast}-\frac{1}{4}F_{\mu \nu}F^{\mu \nu}\Bigr], \label{action}
\end{equation}
where $e$ is the gauge coupling: the charge of the particle, $m$ is the mass of the scalar field, and $\xi$ is a dimensionless nonminimal coupling.
We assume that the complex scalar field is a test field probing two background fields: the gravitational field and the electromagnetic field.
The gravitational field is described by the $\dsf$ metric which reads in the conformal coordinates as
\begin{align}
ds^{2} & =\Omega^{2}(\tau)\Bigl(d\tau^{2}-dx^2-dy^2-dz^2\Bigr), & \tau & \in(-\infty,0), & \x & =(x,y,z)\in\mathbb{R}^{3}, \label{metric}
\end{align}
where the scale factor $\Omega(\tau)$ and the Hubble constant $H$ are given by
\begin{align}
\Omega(\tau) & =-\frac{1}{H \tau}, & H & =\Omega^{-2}(\tau)\frac{d\Omega(\tau)}{d\tau}. \label{hubble}
\end{align}
A $\dsf$ has a scalar curvature $R=12H^2$, therefore the inclusion of a nonminimal coupling term like $\xi R\varphi\varphi^{*}$ would just modify the mass term from $m^{2}$ to $m^{2}+12\xi H^{2}$. For simplicity, we will not consider this in this paper and further set $\xi=0$.
For the electromagnetic field, we consider that it is composed of a constant electric and a constant magnetic part. The vector potential describing the constant electric and the magnetic field parallel to each other in the conformal metric~(\ref{metric}) is given by
\begin{equation}
A_{\mu}(x)=-\frac{E}{H^{2}\tau}\delta_{\mu}^{3}+By\delta_{\mu}^{1}, \label{potential}
\end{equation}
where $E$ and $B$ are constants. The Klein-Gordon equation then reads from the action~(\ref{action}),
\begin{equation}
\biggl[\rnd_{0}^{2}+2H\Omega(\tau)\rnd_{0}-\bigl(\rnd_{1}+ieBy\bigr)^{2}-\rnd_{2}^{2}
-\Bigl(\rnd_{3}+\frac{ieE}{H}\Omega(\tau)\Bigr)^{2}+m^{2}\Omega^{2}(\tau)\biggr]\varphi(x)=0. \label{kgeq}
\end{equation}
The solution of the spatial part of Eq.~(\ref{kgeq}) is a bit more involved than a simple Fourier transformation because of the explicit $y$-dependence. Substituting
\begin{equation}
\varphi(x)=\Omega^{-1}(\tau)\tilde{\varphi}(x), \label{tilde}
\end{equation}
into Eq.~(\ref{kgeq}) yields
\begin{equation}
\biggl[\rnd_{0}^{2}-\bigl(\rnd_{1}+ieBy\bigr)^{2}-\rnd_{2}^{2}-\Bigl(\rnd_{3}+\frac{ieE}{H}\Omega(\tau)\Bigr)^{2}
+m^{2}\Omega^{2}(\tau)-2H^{2}\Omega^{2}(\tau)\biggr]\tilde{\varphi}(x)=0. \label{tildeq}
\end{equation}
Using the ansatz
\begin{equation}
\tilde{\varphi}(x)=e^{\pm i\x\cdot\ky}h^{\pm}(y)f^{\pm}(\tau), \label{ansatz}
\end{equation}
where we have defined
\begin{equation}
\ky:=(k_{x},0,k_{z}), \label{ky}
\end{equation}
and $\pm$ denotes the positive and negative frequency solutions of Eq.~(\ref{tildeq}), respectively, we decouple the spatial and time dependent parts of Eq.~(\ref{tildeq}) as
\begin{eqnarray}
\frac{d^{2}h^{\pm}(y)}{dy^{2}}-\Bigl(eBy\pm k_{x}\Bigr)^{2}h^{\pm}(y)=-sh^{\pm}(y), \label{ypart} \\
\frac{d^{2}f^{\pm}(\tau)}{d\tau^{2}}+\biggl[\Bigl(\frac{eE}{H^{2}\tau}\mp k_{z}\Bigr)^{2}+\frac{m^{2}}{H^{2}\tau^{2}}
-\frac{2}{\tau^{2}}\biggr]f^{\pm}(\tau)=-sf^{\pm}(\tau). \label{tpart}
\end{eqnarray}
The harmonic wave function $h^{\pm}(y)$ is a Landau state given by
\begin{align}
h_{n}(y_{\pm}) & =\sqrt{\frac{\sqrt{eB}}{\sqrt{\pi}2^{n} n!}}\exp \Bigl(-\frac{y_{\pm}^{2}}{2}\Bigr)H_{n}(y_{\pm} ),
& y_{\pm} & :=\sqrt{eB}y\pm \frac{k_x}{\sqrt{eB}}, \label{hermite}
\end{align}
where $H_{n}$ with $n\in \mathbb{N}$ is the Hermite polynomial and $s$ is the Landau energy
\begin{equation}
s=(2n+1)eB. \label{s}
\end{equation}
The normalized wave functions~(\ref{hermite}) satisfy the orthonormality relation
\begin{equation}
\int_{-\infty}^{+\infty}dy h_{n}(y_{\pm})h_{n'}(y_{\pm})=\delta_{n,n'}, \label{h:orthonor}
\end{equation}
and the completeness relation
\begin{equation}
\sum_{n=0}^{\infty}h_{n}(y_{\pm}) h_{n}(y_{\pm}')=\delta(y-y'), \label{h:completn}
\end{equation}
where $y_{\pm}'$ is given by replacing $y$ by $y'$ in the definition~(\ref{hermite}) of $y_{\pm}$. We note that the standard prescription in a flat
spacetime applies also to our results; when one adds a magnetic field, the pair creation in the general case can be deduced from the pure electric field
case $(B=0)$ by replacing the transverse momentum squared $\bf{k}_{\perp}^{2}$ by the Landau levels $(2n+1)eB$.
Following Refs.~\cite{Bavarsad:2016cxh,Kim:2016dmm}, we find the positive and negative frequency solutions with desired asymptotic forms at early
times ($\tau\rightarrow -\infty$), i.e., the in-vacuum mode functions are given by the Hadamard states
\begin{eqnarray}
U_{\vin}\bigl(x;\ky,n\bigr) &=& \frac{e^{\frac{i\pi\kappa}{2}}}{\sqrt{2k}}\Omega^{-1}(\tau)e^{+i\x\cdot\ky}h_{n}(y_{+})
W_{\kappa,\gamma}\bigl(e^{\frac{-i\pi}{2}}2p\bigr), \label{uin} \\
V_{\vin}\bigl(x;\ky,n\bigr) &=& \frac{e^{-\frac{i\pi\kappa}{2}}}{\sqrt{2k}}\Omega^{-1}(\tau)e^{-i\x\cdot\ky}h_{n}(y_{-})
W_{\kappa,-\gamma}\bigl(e^{\frac{+i\pi}{2}}2p\bigr). \label{vin}
\end{eqnarray}
Similarly, the positive and negative frequency solutions with desired asymptotic forms at late times ($\tau\rightarrow 0$), i.e., the out-vacuum mode functions are given by
\begin{eqnarray}
U_{\vout}\bigl(x;\ky,n\bigr) &=& \frac{e^{\frac{i\pi\gamma}{2}}}{\sqrt{4|\gamma|k}}\Omega^{-1}(\tau)e^{+i\x\cdot\ky}h_{n}(y_{+})
M_{\kappa,\gamma}\bigl(e^{\frac{-i\pi}{2}}2p\bigr), \label{uout} \\
V_{\vout}\bigl(x;\ky,n\bigr) &=& \frac{e^{\frac{i\pi\gamma}{2}}}{\sqrt{4|\gamma|k}}\Omega^{-1}(\tau)e^{-i\x\cdot\ky}h_{n}(y_{-})
M_{\kappa,-\gamma}\bigl(e^{\frac{+i\pi}{2}}2p\bigr). \label{vout}
\end{eqnarray}
Here, $W_{\kappa,\gamma}$ and $M_{\kappa,\gamma}$ are some hypergeometrical functions known as the Whittaker functions \cite{Book:Nist} and the parameters have been defined as
\begin{align}
k & =\sqrt{k_{z}^{2}+(2n+1)eB}, & r & =\frac{k_{z}}{k}, & p & =-\tau k, \nn \\
\py & =-\tau\ky, & \ell & =eB\tau^{2}, & \mu & =\frac{m}{H}, \nn \\
\lambda & =\frac{eE}{H^{2}}, & \kappa & =i\lambda r, & \gamma & =\sqrt{\frac{9}{4}-\lambda^{2}-\mu^{2}}. \label{parameters}
\end{align}
In Secs.~\ref{sec:KG} and~\ref{sec:Schwinger} of this paper, we assume the \textit{semiclassical condition},
\begin{equation}
\lambda^{2}+\mu^{2}\gg 1, \label{semi}
\end{equation}
hence the parameter $\gamma$ is purely imaginary. We adopt the sign convention $\gamma=+i|\gamma|$.
\par

The orthonormality relations
\begin{eqnarray}
\Bigl(U_{\vin(\vout)}(x;\ky,n),U_{{\vin(\vout)}}(x;\ky',n')\Bigr) &=&
- \Bigl(V_{\vin(\vout)}(x;\ky,n),V_{{\vin(\vout)}}(x;\ky',n')\Bigr) \nn \\
&=& (2\pi)^{2}\delta^{2}(\ky-\ky')\delta_{n,n'}, \nn \\
\Bigl(U_{\vin(\vout)}(x;\ky, n),V_{{\vin(\vout)}}(x;\ky',n')\Bigr) &=& 0, \label{orthonor}
\end{eqnarray}
can be shown to hold. Using two complete sets of orthonormal mode functions, we expand the scalar field operator.
In terms of the in-mode functions we can express the field operator as
\begin{equation}
\varphi(x)=\sum_{n=0}^{\infty}\int\frac{d^{2}\dky}{(2\pi)^{2}}\Bigl[U_{\vin}\bigl(x;\ky,n\bigr)a_{\vin}(\ky,n)
+V_{\vin}\bigl(x;\ky,n\bigr)b_{\vin}^{\dag}(\ky,n)\Bigr], \label{phiin}
\end{equation}
where the operator $a_{\vin}$ annihilates a particle and the operator $b_{\vin}^{\dag}$ creates an antiparticle in the state with the momentum $\ky$ and
the Landau level $n$. The quantization is implemented by imposing the commutation relations
\begin{equation}
\Bigl[a_{\vin}(\ky,n),a_{\vin}^{\dag}(\ky',n')\Bigr] = \Bigl[b_{\vin}(\ky,n),b_{\vin}^{\dag}(\ky',n')\Bigr]
=(2\pi)^{2}\delta^{2}(\ky-\ky') \delta_{n,n'}, \label{commin}
\end{equation}
and the in-vacuum state is defined as
\begin{align}
a_{\vin}(\ky,n)\big|\vin\big\rangle & =0, & \forall \ky, n. \label{vacin}
\end{align}
We can expand the scalar field operator in terms of the out-mode functions and we similarly define the out-annihilation $a_{\vout}$ and creation
$b_{\vout}^{\dag}$ operators as
\begin{equation}
\varphi(x)=\sum_{n=0}^{\infty}\int\frac{d^{2}\dky}{(2\pi)^{2}}\Bigl[U_{\vout}\bigl(x;\ky,n\bigr)a_{\vout}(\ky,n)+V_{\vout}
\bigl(x;\ky,n\bigr)b_{\vout}^{\dag}(\ky,n)\Bigr], \label{phiout}
\end{equation}
where the quantization commutation relations are given by
\begin{equation}
\Bigl[a_{\vout}(\ky,n),a_{\vout}^{\dag}(\ky',n')\Bigr] = \Bigl[b_{\vout}(\ky,n),b_{\vout}^{\dag}(\ky',n')\Bigr]
=(2\pi)^{2}\delta^{2}(\ky-\ky')\delta_{n,n'}, \label{commout}
\end{equation}
and the out-vacuum state is defined as
\begin{align}
a_{\vout}(\ky,n)\big|\vout\big\rangle &=0, & \forall \ky, n. \label{vacout}
\end{align}
The canonical momentum $\pi(x)$ conjugated to the scalar field $\varphi(x)$ is defined through the Lagrangian.
It reads from Eq.~(\ref{action}),
\begin{equation}
\pi(x)=\frac{\rnd\mathcal{L}}{\rnd(\rnd_{0}\varphi)}=\Omega^{2}(\tau)\rnd_{0}\varphi^{\ast}. \label{momentum}
\end{equation}
Then, using the explicit form of the scalar field operator $\varphi(x)$ and the canonical momentum $\pi(x)$ in terms of the mode functions, one can verify that the canonical equal-time commutation relation correctly holds
\begin{equation}
\Bigl[\varphi(\tau,\x),\pi(\tau,\x')\Bigr]=i\delta^{3}(\x-\x'). \label{canonic}
\end{equation}
%%%%%%%%%%%%%%%%%%%%%%%%%%%%%%%%%%%%%%%%%%%%%%%%%%%%%%%%%%%%%%%%%%%%%%%%%%%%%%%%%%%%%%%%%%%%%%%%%%%%%%%%%%%%%%%%%%%%%%%%%%%%%%%%%%%%%%%%%%%%%%%%%%%%%%%%%%%%%%%
%%%%%%%%%%%%%%%%%%%%%%%%%%%%%%%%%%%%%%%%%%%%%%%%%%%%%%%%%%%%%%%%%%%%%%%%%%%%%%%%%%%%%%%%%%%%%%%%%%%%%%%%%%%%%%%%%%%%%%%%%%%%%%%%%%%%%%%%%%%%%%%%%%%%%%%%%%%%%%%
\section{\label{sec:Schwinger}Schwinger Effect}
The usual quantity describing the Schwinger effect is the pair creation or decay rate which is derived from the Bogoliubov coefficients
\cite{Book:Parker,Book:Birrell},
\begin{eqnarray}
\mathcal{A}\bigl(\ky,n;\ky',n'\bigr) &=& \Bigl(U_{\vout}\big(x;\ky,n\big),U_{\vin}\big(x;\ky',n'\big)\Bigr), \label{def:alpha} \\
\mathcal{B}\bigl(\ky,n;\ky',n'\bigr) &=& -\Bigl(U_{\vout}\big(x;\ky,n\big),V_{\vin}\big(x;\ky',n'\big)\Bigr). \label{def:beta}
\end{eqnarray}
Substituting the explicit form of the mode functions~(\ref{uin})-(\ref{vout}) into Eqs.~(\ref{def:alpha}) and~(\ref{def:beta}) leads to
\begin{align}
\mathcal{A}\bigl(\ky n;\ky' n'\bigr) & = (2\pi)^{2}\delta^{2}(\ky-\ky')\delta_{n,n'} \alpha, &
\alpha & = \frac{(2|\gamma|)^{\frac{1}{2}}\Gamma\bigl(2\gamma\bigr)}{\Gamma\bigl(\frac{1}{2}+\kappa+\gamma\bigr)}
e^{\frac{i\pi}{2}(\kappa-\gamma)}, \label{alpha} \\
\mathcal{B}\bigl(\ky n;\ky' n'\bigr) &= (2\pi)^{2}\delta^{2}(\ky+\ky')\delta_{n,n'} \beta, &
\beta &=-i\frac{(2|\gamma|)^{\frac{1}{2}}\Gamma\bigl(-2\gamma\bigr)}{\Gamma\bigl(\frac{1}{2}+\kappa-\gamma\bigr)}
e^{\frac{i\pi}{2}(\kappa+\gamma)}, \label{beta}
\end{align}
where the coefficients satisfy the bosonic relation $|\alpha|^{2}-|\beta|^{2}=1$.
A Bogoliubov transformation relates the out-operator $a_{\vout}$ to the in-operator $a_{\vin}$ as
\begin{eqnarray}
a_{\vout}(\ky,n) = \sum_{n'=0}^{\infty}\int\frac{d^{2}\dky'}{(2\pi)^{2}}\Bigl[\mathcal{A}^{\ast}\bigl(\ky,n;\ky',n'\bigr)a_{\vin}(\ky',n')
- \mathcal{B}^{\ast}\bigl(\ky,n;\ky',n'\bigr)b_{\vin}^{\dag}(\ky',n')\Bigr]. \label{transform}
\end{eqnarray}
Using the out-operator $a_{\vout}(\ky,n)$, we can calculate the expected number of the created pairs with the comoving momentum $\ky$ and the Landau level
$n$ carried by the in-vacuum state
\begin{equation}
\frac{1}{L_{x}L_{z}}\Big\langle \vin\Bigl| a_{\vout}^{\dag}(\ky,n)a_{\vout}(\ky,n)\Bigr|\vin \Big\rangle
=\bigl|\beta\bigl(k_{z},n\bigr)\bigr|^{2}, \label{number}
\end{equation}
where we have used Eqs.~(\ref{beta}),~(\ref{transform}) and, for convenience, the three-volume of the $\dsf$ is normalized into a box with dimensions $V=L_{x}L_{y}L_{z}$. Then the decay rate $\Gamma$, i.e., the number of created pairs $N$ per unit of the physical four-volume of the $\dsf$ is given by
\begin{equation}
\Gamma:=\frac{N}{\sqrt{|g|}TV}=\frac{1}{\Omega^{4}(\tau)TL_{y}}\sum_{n=0}^{\infty}\int\frac{dk_{z}}{(2\pi)}\frac{dk_{x}}{(2\pi)}
\bigl|\beta\bigl(k_{z},n\bigr)\bigr|^{2}, \label{rate}
\end{equation}
where $T$ is the time interval of the pair creation. The Bogoliubov coefficient $\beta$ is independent of the momentum component $k_{x}$ which determines the position of the center of the Gaussian wave packet on $y$ axis by the relation $y=k_{x}/(eB)$.
Consequently, the integral gives \cite{Kuznetsov:2004tb}
\begin{equation}
\int\frac{dk_{x}}{(2\pi)}=\frac{eBL_{y}}{(2\pi)}. \label{kx}
\end{equation}
To perform the $k_{z}$-integral on the right hand side of Eq.~(\ref{rate}), we adopt the semiclassical method used in Refs.~\cite{Frob:2014zka,Kobayashi:2014zza}: most of the particles are created around the time
\begin{equation}
\tau \sim -\frac{|\gamma|}{k_{z}}. \label{relation}
\end{equation}
Imposing the relation~(\ref{relation}) and transforming the $k_{z}$-integral into a $\tau$-integral, we then obtain
\begin{equation}
\Gamma=\frac{H^{4}\ell|\gamma|}{4\pi^{2}}\sum_{n=0}^{\infty}
\frac{e^{2\pi|\kappa|}+e^{-2\pi|\gamma|}}{e^{2\pi|\gamma|}-e^{-2\pi|\gamma|}}, \label{rate:1}
\end{equation}
where
\begin{equation}
|\kappa|=\frac{\lambda|\gamma|}{\sqrt{|\gamma|^{2}+(2n+1)\ell}}. \label{kappa}
\end{equation}
A physical magnetic field in a spatially flat FLRW universe with a cosmological scale factor $\Omega(\tau)$ dilutes as $B\Omega^{-2}(\tau)$ where $B$
behaves as a magnetic field in the comoving spacetime \cite{Tsagas:2007ik,Giovannini:2003yn}.
This preserves the flux conservation for the physical magnetic field. Recalling that $\ell=eB\tau^{2}$, consequently, the decay rate $\Gamma$ depends on
the time $\tau$ due to the dilution of the physical magnetic field. We may write Eq.~(\ref{rate:1}) in another form
\begin{equation}
\Gamma=\Bigl(\frac{eB\Omega^{-2}}{2\pi}\Bigr)\Bigl(\frac{H^{2}|\gamma|}{2\pi}\Bigr)
\sum_{n=0}^{\infty}\biggl[\frac{e^{2\pi|\kappa|}-1}{e^{2\pi|\gamma|}-e^{-2\pi|\gamma|}}
+\frac{1}{e^{2\pi|\gamma|}-1}\biggr]. \label{rate:2}
\end{equation}
The first term in the square bracket in Eq.~(\ref{rate:2}) is the pair creation rate from the electromagnetic field while the second term is the dS
radiation with a new temperature $T=m/(2\pi|\gamma|)$ weighted by the density of states for the electromagnetic field.
\par

A few comments are in order.
First, there is a term independent of the Landau levels, whose sum apparently gives a diverging factor. We tackle this issue by using the Riemann zeta function prescription as in Ref.~\cite{Elizalde:1994gf}. We also use the $n=0$ term which gives a constant factor
\begin{equation}
\sum_{n=0}^{\infty}=1+\zeta(0)=\frac{1}{2}, \label{zeta}
\end{equation}
where Eq.~(\ref{special}) has been used. Thus, the pair production from the zeta regularization technique leads to a finite result
\begin{equation}
\Gamma=\Bigl(\frac{eB\Omega^{-2}}{2\pi}\Bigr)\Bigl(\frac{H^{2}|\gamma|}{2\pi}\Bigr)\Bigl(\frac{1}{e^{4\pi|\gamma|}-1}\Bigr)
\biggl[\frac{1}{2}+\sum_{n=0}^{\infty}e^{2\pi(|\kappa|+|\gamma|)}\biggr]. \label{rate:3}
\end{equation}
Second, in the regime of the weak magnetic field: $\ell\ll\min(1,\mu,\lambda)$ and the strong electric field: $\lambda\gg\max(1,\mu,\ell)$, Eq.~(\ref{rate:3}) leads to
\begin{equation}
\Gamma=\frac{1}{2}\Bigl(\frac{eB\Omega^{-2}}{2\pi}\Bigr)\Bigl(\frac{eE}{2\pi}\Bigr)e^{\frac{-\pi m^{2}}{|eE|}}. \label{rate:4}
\end{equation}
Third, in the limit of zero electric field $E=0$, the first term in the square bracket of Eq.~(\ref{rate:2}) vanishes and the second term is the dS
radiation with the Gibbons-Hawking temperature \cite{Gibbons:1977mu}
\begin{equation}
\Gamma=\frac{1}{2}\Bigl(\frac{eB\Omega^{-2}}{2\pi}\Bigr)\Bigl(\frac{H^{2}|\gamma|}{2\pi}\Bigr)\frac{1}{e^{2\pi|\gamma|}-1}. \label{rate:5}
\end{equation}
The factor $1/2$ comes from the spin multiplicity for spinless bosons while it is $1$ for spin $1/2$ fermions.
The radiation in the pure $\dsf$ without electromagnetic fields consists of massive particles $m\geq 3H/2$ and the leading term of $H^2|\gamma|$ is $Hm$
for the density of states \cite{Kim:2010cb}. Thus, the presence of a cosmic magnetic field enhances the dS radiation through the density of states by a factor of $eB\Omega^{-2}$. The density of states $eB$ becomes $H^{2}$ when there is no magnetic field.
Finally, in the Minkowski spacetime limit $H=0$, Eq.~(\ref{rate:1}) gives the Schwinger formula in scalar QED \cite{Schwinger:1951nm}
\begin{equation}
\Gamma=\frac{1}{2}\Bigl(\frac{eB}{2\pi}\Bigr)\Bigl(\frac{eE}{2\pi}\Bigr)\frac{e^{\frac{-\pi m^{2}}{|eE|}}}{\sinh\Bigl(\frac{\pi B}{E}\Bigr)}.
\label{rate:6}
\end{equation}
%%%%%%%%%%%%%%%%%%%%%%%%%%%%%%%%%%%%%%%%%%%%%%%%%%%%%%%%%%%%%%%%%%%%%%%%%%%%%%%%%%%%%%%%%%%%%%%%%%%%%%%%%%%%%%%%%%%%%%%%%%%%%%%%%%%%%%%%%%%%%%%%%%%%%%%%%%%%%%%
%%%%%%%%%%%%%%%%%%%%%%%%%%%%%%%%%%%%%%%%%%%%%%%%%%%%%%%%%%%%%%%%%%%%%%%%%%%%%%%%%%%%%%%%%%%%%%%%%%%%%%%%%%%%%%%%%%%%%%%%%%%%%%%%%%%%%%%%%%%%%%%%%%%%%%%%%%%%%%%
\section{\label{sec:current}Induced Current}
Semiclassically, the conductive current $J_{\sem}$ of the newly created Schwinger pairs having a charge $e$, a number density $\mathcal{N}$, and a
velocity $v$ due to the background electric field is defined as $J_{\sem}=2e\mathcal{N}v$.
The number density of the semiclassical Schwinger pairs at the time $\tau$ reads
\begin{equation}
\mathcal{N}(\tau)=\Omega^{-2}(\tau)\int_{0}^{\tau}\Omega^{4}(\tau')\Gamma(\tau') d\tau' \sim \frac{\Gamma(\tau)}{H}, \label{mathN}
\end{equation}
where $\Gamma$ is given by Eq.~(\ref{rate:1}). The current $J_{\sem}$ is valid under the semiclassical condition, which is given by Eq.~(\ref{semi}).
In this section we investigate the in-vacuum expectation value of the current operator which is referred to as the induced current, without assuming the constraint~(\ref{semi}) on the parameters.
Hence, $\gamma$ can be real or purely imaginary depending on the value of involved parameters, $\lambda$ and $\mu$.
\par

\begin{figure}[t]
\centering
\includegraphics[scale=0.7]{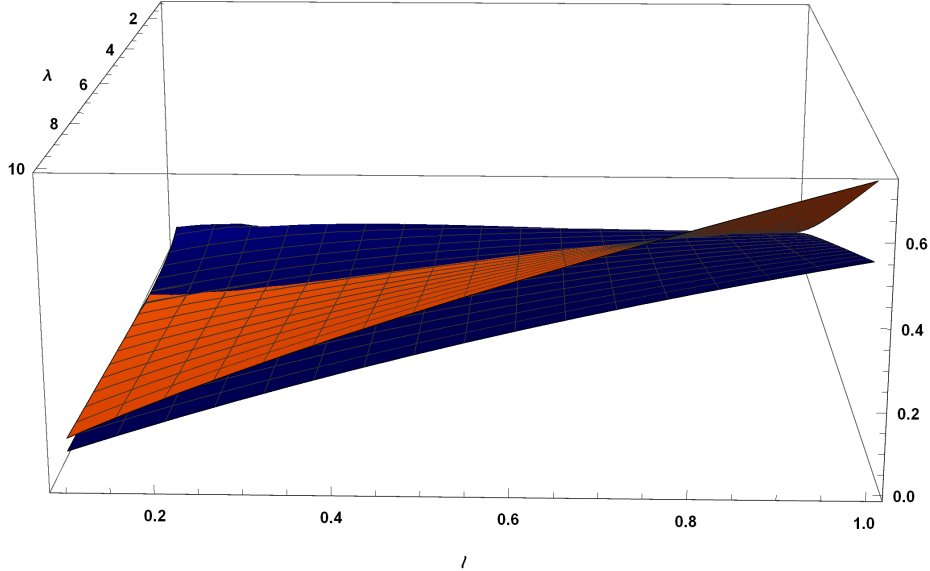}
\caption{The normalized induced current $J/eH^{3}$ (upper surface) and semiclassical current $J_{\sem}/eH^{3}$ (lower surface) are plotted as functions
of $\lambda$ and $\ell$, in the lowest Landau state $n=0$ with $\mu=1$.} \label{fig:1}
\end{figure}

\begin{figure}[t]
\centering
\includegraphics[scale=0.7]{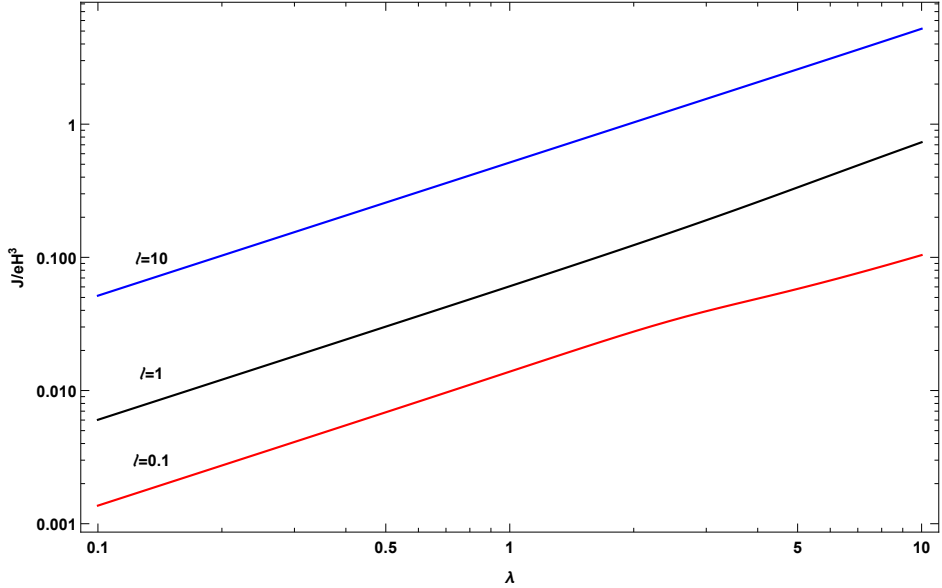}
\caption{For different values of $\ell$, the normalized induced current $J/eH^{3}$ is plotted as a function of $\lambda$,
in the lowest Landau state $n=0$ with $\mu=1$.} \label{fig:2}
\end{figure}

\begin{figure}[t]
\centering
\includegraphics[scale=0.7]{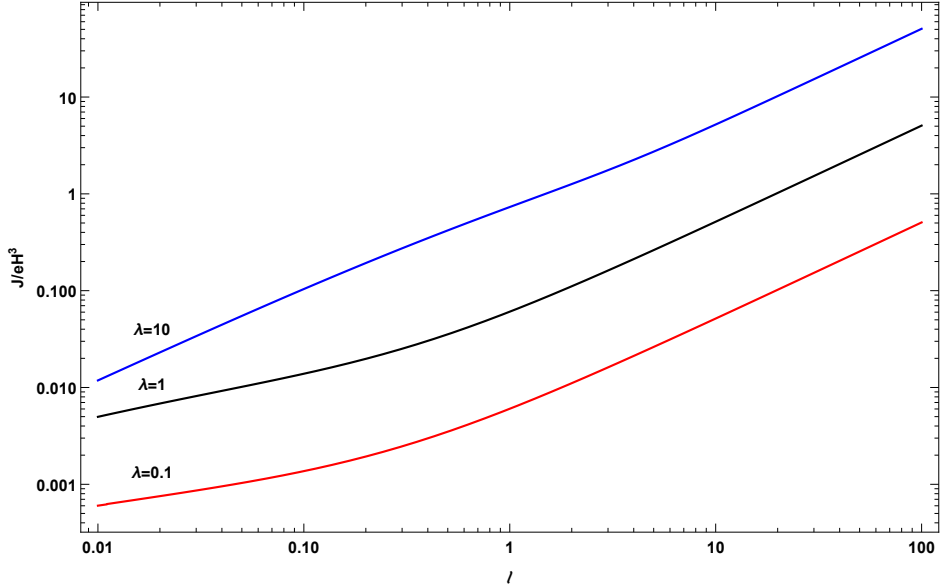}
\caption{For different values of $\lambda$, the normalized induced current $J/eH^{3}$ is plotted as a function of $\ell$,
in the lowest Landau state $n=0$ with $\mu=1$.} \label{fig:3}
\end{figure}

The current operator is defined by
\begin{equation}
j^{\mu}(x)=\frac{ie}{2}g^{\mu\nu}\Bigl(\bigl\{\bigl(\rnd_{\nu}+ieA_{\nu}\bigr)\varphi,\varphi^{*}\bigr\}
-\bigl\{\bigl(\rnd_{\nu}-ieA_{\nu}\bigr)\varphi^{*},\varphi\bigr\}\Bigr), \label{current}
\end{equation}
and can be shown to be conserved $\nabla_{\mu}j^{\mu}=0$ \cite{Book:Parker}. In order to compute the expectation value of the current operator, we will
use the in-vacuum state since it is a Hadamard state \cite{Frob:2014zka,Garriga:1994bm}. Substituting the scalar field operator~(\ref{phiin}) into the current expression~(\ref{current}) and using Eqs.~(\ref{commin}) and~(\ref{vacin}), we find that the only nonvanishing component of the current is the
one parallel to the electric field which is given by
\begin{eqnarray}
\bigl\langle\vin\big|j^{3}(x)\big|\vin\bigr\rangle &=& \frac{eH^{2}}{4\pi^{2}}\Omega^{-2}(\tau)\sum_{n=0}^{\infty}
\int_{-\infty}^{+\infty}\frac{dp_{z}}{p}\big(rp+\lambda\big)e^{-\pi\lambda r}\big|W_{i\lambda r,\gamma}(-2ip)\big|^{2} \nn \\
&\times & \int_{-\infty}^{+\infty}dp_{x}h_{n}^{2}(y_{+}). \label{jz}
\end{eqnarray}
Using the orthonormality relation~(\ref{h:orthonor}) the $p_{x}$-integral is performed
\begin{equation}
\int_{-\infty}^{+\infty}dp_{x}h_{n}^{2}(y_{+})=-eB\tau. \label{intpx}
\end{equation}
If we parameterize the induced current as
\begin{equation}
J=\Omega(\tau)\bigl\langle\vin\big|j^{3}(x)\big|\vin\bigr\rangle, \label{par}
\end{equation}
then Eq.~(\ref{jz}) is simplified to
\begin{equation}
J=\frac{eH^{3}\ell}{4\pi^{2}}\sum_{n=0}^{\infty}\int_{-\infty}^{+\infty}\frac{dp_{z}}{p}\big(rp+\lambda\big)e^{-\pi\lambda r}
\big|W_{i\lambda r,\gamma}(-2ip)\big|^{2}. \label{j}
\end{equation}
The remaining integral in the induced current expression~(\ref{j}) deals with the Whittaker functions. In the absence of the magnetic field background,
the translational symmetry helps performing the integral using the Mellin-Barnes representation of the Whittaker functions;
see \cite{Frob:2014zka,Kobayashi:2014zza}.
However, even in this case the exact expression for the induced current is very complicated and one has to look at limiting regimes to better grasp the physics of the results.
In the regime of $\lambda\gg 1$ the semiclassical condition~(\ref{semi}) is satisfied, and the induced current~(\ref{j}) is comparable to the
semiclassical current $J_{\sem}=2e\mathcal{N}v$. Considering the ultrarelativistic particles with velocity $v\sim 1$, Fig.~\ref{fig:1} shows that the induced current $J$ approaches the semiclassical current $J_{\sem}$ for the strong electric field regime $\lambda\gg\max(1,\mu,\ell)$.
In Figs.~\ref{fig:2} and~\ref{fig:3} we plot the induced current expression~(\ref{j}) as a function of the electric and magnetic fields, respectively.
The figures illustrate that the induced current of a massive scalar field responds to the strong electromagnetic field as $J\varpropto B\cdot E$; for additional numerical investigations see \cite{Proceeding}. As a matter of consistency, we will now analytically investigate the limiting behavior of the induced current~(\ref{j}) to show that it agrees with the numerical investigations.
%%%%%%%%%%%%%%%%%%%%%%%%%%%%%%%%%%%%%%%%%%%%%%%%%%%%%%%%%%%%%%%%%%%%%%%%%%%%%%%%%%%%%%%%%%%%%%%%%%%%%%%%%%%%%%%%%%%%%%%%%%%%%%%%%%%%%%%%%%%%%%%%%%%%%%%%%%%%%%%
%%%%%%%%%%%%%%%%%%%%%%%%%%%%%%%%%%%%%%%%%%%%%%%%%%%%%%%%%%%%%%%%%%%%%%%%%%%%%%%%%%%%%%%%%%%%%%%%%%%%%%%%%%%%%%%%%%%%%%%%%%%%%%%%%%%%%%%%%%%%%%%%%%%%%%%%%%%%%%%
\subsection{\label{sec:weak}Weak magnetic field regime}
In the weak magnetic field regime the relation $\ell\ll\min(1,\mu,\lambda)$ is satisfied.
Taking the limit $\ell\rightarrow 0$ in the momentum $p$ gives $p\sim|p_{z}|$; see definition of $p$ in Eq.~(\ref{parameters}).
Then the induced current expression~(\ref{j}) is simplified to
\begin{equation}
J\simeq\frac{eH^{3}\ell}{4\pi^{2}}\sum_{n=0}^{\infty}\sum_{r=\pm 1}\int_{0}^{\infty}\frac{dp_{z}}{p_{z}}\big(rp_{z}+\lambda\big)e^{-\pi\lambda r}
\big|W_{i\lambda r,\gamma} \big(-2ip_{z}\big)\big|^{2}. \label{jweak}
\end{equation}
The integrand in the right hand side of Eq.~(\ref{jweak}) is independent of the Landau states.
Hence, similarly to the prescription used in Sec.~\ref{sec:Schwinger}, using the zeta function representation~(\ref{zeta}), the current
expression~(\ref{jweak}) is regularized to
\begin{equation}
J\simeq\frac{eH^{3}\ell}{8\pi^{2}}\sum_{r=\pm 1}\int_{0}^{\infty}\frac{dp_{z}}{p_{z}}\big(rp_{z}+\lambda\big)e^{-\pi\lambda r}
\big|W_{i\lambda r,\gamma} \big(-2ip_{z}\big)\big|^{2}. \label{jreg}
\end{equation}
The computation and adiabatic regularization of the current~(\ref{jreg}) have been reviewed in Appendix~\ref{app:reg} and the final result can be read
from Eq.~(\ref{scheme}). We then obtain
\begin{equation}
J_{\reg}=\Bigl(\frac{eH^{3}}{4\pi^{2}}\Bigr)\frac{\ell\gamma\sinh\big(2\pi\lambda\big)}{\sin\big(2\pi\gamma\big)}. \label{jregweak}
\end{equation}
We comment here that our result is unlike the case of a pure electric field in $\dsf$ \cite{Kobayashi:2014zza}, where in order to renormalize the current
an adiabatic expansion up to order two has been performed to remove the quadratic divergence, here the adiabatic order zero is enough as in the $\dst$
case \cite{Frob:2014zka}. The reason is that we deal here with an effective integration in 1+1 dimensions, and the integration over momentum in the directions orthogonal to the magnetic field is replaced by a discrete sum over quantized Landau levels, which is regularized and renormalized by using
the Riemann zeta function technique; see Appendix~\ref{app:zeta}.
%%%%%%%%%%%%%%%%%%%%%%%%%%%%%%%%%%%%%%%%%%%%%%%%%%%%%%%%%%%%%%%%%%%%%%%%%%%%%%%%%%%%%%%%%%%%%%%%%%%%%%%%%%%%%%%%%%%%%%%%%%%%%%%%%%%%%%%%%%%%%%%%%%%%%%%%%%%%%%%
%%%%%%%%%%%%%%%%%%%%%%%%%%%%%%%%%%%%%%%%%%%%%%%%%%%%%%%%%%%%%%%%%%%%%%%%%%%%%%%%%%%%%%%%%%%%%%%%%%%%%%%%%%%%%%%%%%%%%%%%%%%%%%%%%%%%%%%%%%%%%%%%%%%%%%%%%%%%%%%
\subsubsection{\label{sec:large}Strong electric field regime}
In the strong electric field regime the relation $\lambda\gg\max(1,\mu,\ell)$ is satisfied.
Taking the limit $\lambda\rightarrow\infty$ in the regularized induced current~(\ref{jregweak}) with $\mu$ and $\ell$ fixed, to the leading order term, gives rise to
\begin{equation}
J_{\reg}\simeq\frac{e}{H}\Bigl(\frac{eB\Omega^{-2}}{2\pi}\Bigr)\Bigl(\frac{eE}{2\pi}\Bigr)e^{\frac{-\pi m^{2}}{|eE|}}. \label{stronge}
\end{equation}
In this regime the decay rate is given by Eq.~(\ref{rate:4}) and does the semiclassical current follow from Eq.~(\ref{mathN}).
Then, one can verify that the induced current~(\ref{stronge}) agrees with the semiclassical current for particles with the velocity $v\sim 1$. %%%%%%%%%%%%%%%%%%%%%%%%%%%%%%%%%%%%%%%%%%%%%%%%%%%%%%%%%%%%%%%%%%%%%%%%%%%%%%%%%%%%%%%%%%%%%%%%%%%%%%%%%%%%%%%%%%%%%%%%%%%%%%%%%%%%%%%%%%%%%%%%%%%%%%%%%%%%%%%
%%%%%%%%%%%%%%%%%%%%%%%%%%%%%%%%%%%%%%%%%%%%%%%%%%%%%%%%%%%%%%%%%%%%%%%%%%%%%%%%%%%%%%%%%%%%%%%%%%%%%%%%%%%%%%%%%%%%%%%%%%%%%%%%%%%%%%%%%%%%%%%%%%%%%%%%%%%%%%%
\subsubsection{\label{sec:heavy}Weak electric field and heavy scalar field regime}
In this regime the relations $\lambda\ll 1$ and $\mu\gg 1$ are satisfied.
Taking the limits $\lambda\rightarrow 0$ and $\mu\rightarrow\infty$ in the regularized induced current expression~(\ref{jregweak}) with $\ell$ fixed,
to the leading order, gives rise to
\begin{equation}
J_{\reg}\simeq\frac{4\pi em}{H^{2}}\Bigl(\frac{eB\Omega^{-2}}{2\pi}\Bigr)\Bigl(\frac{eE}{2\pi}\Bigr)e^{\frac{-2\pi m}{H}}. \label{weake}
\end{equation}
In this regime the decay rate reads from Eq.~(\ref{rate:5}) and the induced current~(\ref{weake}) agrees with the semiclassical current $J_{\sem}$ for particles with the velocity $v\sim (4\pi eE)/H^{2}$.
%%%%%%%%%%%%%%%%%%%%%%%%%%%%%%%%%%%%%%%%%%%%%%%%%%%%%%%%%%%%%%%%%%%%%%%%%%%%%%%%%%%%%%%%%%%%%%%%%%%%%%%%%%%%%%%%%%%%%%%%%%%%%%%%%%%%%%%%%%%%%%%%%%%%%%%%%%%%%%%
%%%%%%%%%%%%%%%%%%%%%%%%%%%%%%%%%%%%%%%%%%%%%%%%%%%%%%%%%%%%%%%%%%%%%%%%%%%%%%%%%%%%%%%%%%%%%%%%%%%%%%%%%%%%%%%%%%%%%%%%%%%%%%%%%%%%%%%%%%%%%%%%%%%%%%%%%%%%%%%
\subsubsection{\label{sec:irhc}Infrared regime}
In this regime the relations $\ell\ll\mu\ll\lambda\ll 1$ are satisfied. Hence the semiclassical current cannot be compared to the induced current in this regime. Taking the limits $\lambda\rightarrow 0$ and $\mu\rightarrow 0$ in the induced current expression~(\ref{jregweak}), we then find
\begin{equation}
J_{\reg}\simeq\frac{9eH^{3}}{8\pi^{2}}\Bigl(\frac{\ell\lambda}{\lambda^{2}+\mu^{2}}\Bigr), \label{irhc:1}
\end{equation}
or in terms of dimensionful variables
\begin{equation}
J_{\reg}\simeq\frac{9eH^{3}}{2}\Bigl(\frac{eB\Omega^{-2}}{2\pi}\Big)\Bigl(\frac{eE}{2\pi}\Big)
\biggl(\frac{1}{\bigl(eE\bigr)^{2}+\bigl(mH\bigr)^{2}}\biggr). \label{irhc:2}
\end{equation}
In this regime for an interval of $\mu\lesssim\lambda\lesssim 1$ a decreasing electric field gives rise to an increasing current and consequently
hyperconductivity. This infrared phenomenon was first discovered in \cite{Frob:2014zka} and dubbed as the infrared hyperconductivity (IRHC) for the case
of a scalar field coupled to a constant, purely electric field background in $\dst$.
In Ref.~\cite{Kobayashi:2014zza}, using an alternative approach in Ref.~\cite{Geng:2017zad}, the authors have computed the current due to a pure electric field in $\dsf$ and found the IRHC. In \cite{Kobayashi:2014zza}, the second order adiabatic expansion leads to a term of the form $\log(m/H)$ in the regularized induced current expression. Therefore, it was not possible to discuses IRHC for the case of a massless minimally coupled scalar field.
However, we note here that the inclusion of the magnetic field and the change of the renormalization prescription allow to explore IRHC in the massless limit. We find indeed that the induced current responds as $J\sim B/E$ and increases unboundedly in the case of a massless minimally coupled scalar field.
For a massive scalar field, an upper bound on the induced current occurs at $\lambda=\mu$ and is given by
\begin{equation}
J_{\reg}\simeq\frac{9eH^{2}}{8\pi m}\Bigl(\frac{eB\Omega^{-2}}{2\pi}\Big). \label{bound}
\end{equation}
The exact nature of IRHC remains a mystery but has been reported in various works in the previous years. It is unexpected as for a decreasing cause: the electric field background, the consequence: the induced current due to the creation of Schwinger pairs increases. It might be a signal for the need for backreaction and the breaking of the working assumptions of the toy model used to derive it or could have a deeper physical meaning that is to be understood. In any case, if it is confirmed within the next years, it has to be taken into account and will give constraints on inflation scenarios.
%%%%%%%%%%%%%%%%%%%%%%%%%%%%%%%%%%%%%%%%%%%%%%%%%%%%%%%%%%%%%%%%%%%%%%%%%%%%%%%%%%%%%%%%%%%%%%%%%%%%%%%%%%%%%%%%%%%%%%%%%%%%%%%%%%%%%%%%%%%%%%%%%%%%%%%%%%%%%%%
%%%%%%%%%%%%%%%%%%%%%%%%%%%%%%%%%%%%%%%%%%%%%%%%%%%%%%%%%%%%%%%%%%%%%%%%%%%%%%%%%%%%%%%%%%%%%%%%%%%%%%%%%%%%%%%%%%%%%%%%%%%%%%%%%%%%%%%%%%%%%%%%%%%%%%%%%%%%%%%
\subsection{\label{sec:strong}Strong magnetic field regime}
In the strong magnetic field regime the relation $\ell\gg\max(1,\mu,\lambda)$ is satisfied.
In this regime, in order to examine the limiting behaviour of the induced current, it is convenient to rewrite Eq.~(\ref{j}) in the form of
\begin{equation}
J=\frac{eH^{3}\ell}{4\pi^{2}}\sum_{n=0}^{\infty}\int_{-1}^{+1}\frac{dr}{(1-r^{2})}
\bigl(rp(r)+\lambda\bigr)e^{-\pi\lambda r}\Bigl|W_{i\lambda r,\gamma}\bigl(-2ip(r)\bigr)\Bigr|^{2}, \label{jr}
\end{equation}
where the momentum $p$ as a function of $r$ is given by
\begin{equation}
p(r)=\sqrt{\frac{(1+2n)\ell}{1-r^{2}}}. \label{pr}
\end{equation}
In the limit of $\ell\rightarrow\infty$ and as a consequence $p(r)\rightarrow\infty$, the Whittaker function approximates
\begin{equation}
\Bigl|W_{i\lambda r,\gamma}\bigl(-2ip(r)\bigr)\Bigr|^{2} \sim e^{\pi\lambda r}. \label{form}
\end{equation}
Substituting the asymptotic form~(\ref{form}) into Eq.~(\ref{jr}) and using the prescription~(\ref{zeta}), we obtain
\begin{equation}
J \simeq \frac{eH^{3}\ell\lambda}{8\pi^{2}}\int_{-1}^{+1}\frac{dr}{(1-r^{2})}. \label{jstrong}
\end{equation}
In order to regularize the integral in Eq.~(\ref{jstrong}), we use following prescription
\begin{equation}
\int_{-1}^{+1} \frac{dr}{(1-r^{2})} = \sum_{n=0}^{\infty}\int_{-1}^{+1}drr^{2n}
= \sum_{n=0}^{\infty}\frac{1}{n+\frac{1}{2}}, \label{intr}
\end{equation}
and using the definition of the Hurwitz zeta function given by Eq.~(\ref{def:Hurwitz}), we represent the summation as
\begin{equation}
\sum_{n=0}^{\infty} \frac{1}{n+\frac{1}{2}} =
-\frac{\rnd^{2}}{\rnd a\rnd s}\zeta\Bigl(s=0,a=\frac{1}{2}\Bigr). \label{sum}
\end{equation}
Finally, with Eqs.~(\ref{jstrong})-(\ref{sum}) and~(\ref{formula}), we obtain the regularized induced current in the strong magnetic field regime
\begin{eqnarray}
J_{\reg} \simeq \Big(\gamma_{\mathrm{Euler}}+\log(4)\Big)\frac{eH^{3}\ell\lambda}{8\pi^{2}}
\sim \frac{e}{H}\Big(\frac{eB\Omega^{-2}}{2\pi}\Big)\Big(\frac{eE}{2\pi}\Big), \label{strong}
\end{eqnarray}
where $\gamma_{\mathrm{Euler}}=0.577\cdots$ is the Euler's constant.
This result shows the new contribution of the magnetic field in the strong magnetic field regime.
As for the strong electric field regimes, the induced current presents a linear behavior in the magnetic field.
As expected, it is the pair production due to the electromagnetic field which dominates its gravitational counterpart, in this regime.
%%%%%%%%%%%%%%%%%%%%%%%%%%%%%%%%%%%%%%%%%%%%%%%%%%%%%%%%%%%%%%%%%%%%%%%%%%%%%%%%%%%%%%%%%%%%%%%%%%%%%%%%%%%%%%%%%%%%%%%%%%%%%%%%%%%%%%%%%%%%%%%%%%%%%%%%%%%%%%%
%%%%%%%%%%%%%%%%%%%%%%%%%%%%%%%%%%%%%%%%%%%%%%%%%%%%%%%%%%%%%%%%%%%%%%%%%%%%%%%%%%%%%%%%%%%%%%%%%%%%%%%%%%%%%%%%%%%%%%%%%%%%%%%%%%%%%%%%%%%%%%%%%%%%%%%%%%%%%%%
\section{\label{sec:concl}Conclusion}
We have investigated for the first time the effect of a uniform magnetic field on the Schwinger pair production and the induced current due to a uniform electric field in $\dsf$. On the one hand, in Minkowski spacetime, a strong constant electric field can create pairs of charged particles from the vacuum
at the cost of electrostatic energy. This is known as the Schwinger effect. A pure magnetic field does not produce any pair of charged particles since the virtual pair from the vacuum immediately annihilates each other. On the other hand, dS can emit radiation of all species of particles. This is known as
the Gibbons-Hawking radiation. Considering those two effects together has been done in the past years. In this case, two important results are that the Gibbons-Hawking radiation enhances the pair production \cite{Cai:2014qba} and the super-horizon behavior of the field leads to a phenomenon of infrared
hyperconductivity for the induced current \cite{Frob:2014zka,Kobayashi:2014zza,Bavarsad:2016cxh,Geng:2017zad}.
\par

In this paper, we add one more ingredient to this setup: we include a uniform magnetic field parallel to the electric field in $\dsf$. The results of
this paper recover the Schwinger effect and the induced current in the absence of a magnetic field, which has been systematically investigated in
Ref.~\cite{Kobayashi:2014zza}. The consequence of a constant magnetic field on the Schwinger effect and the induced current with or without an electric field in $\dsf$ has been extensively studied.
\par

First, the Schwinger effect is enhanced due to the density of states proportional to the magnetic field. Even in the absence of the electric field, the
pair production rate is a product of the Gibbons-Hawking radiation and the magnetic field. This means that a strong magnetic field indeed assists the pair production in dS; see the result in Eq.~(\ref{rate:5}). This is in contrast to the Schwinger effect due to parallel electric and magnetic fields in the Minkowski spacetime, in which the density of states is proportional to both the electric field and the magnetic field and vanishes when the electric field is absent because a pure magnetic field is stable against spontaneous pair production.
\par

Second, the infrared hyperconductivity has been observed in the regime $\mu\ll\lambda\ll 1$, for weak magnetic fields;
see the result in Eq.~(\ref{irhc:2}). This indicates that in the $\ds$: (i) $\mu=m/H\ll 1$, i.e., the Compton wavelength $m^{-1}$ of the charge is much bigger than the Hubble radius $H^{-1}$; (ii) $\lambda=eE/H^{2}\ll1$, i.e., the electric field $E$ is much smaller than the scalar curvature $R=12H^2$;
(iii) $\mu\ll\lambda$ or $eE/H \gg m$, i.e., the electric potential energy across the Hubble radius $H^{-1}$ is much larger than the mass of charge.
This is in contrast to the regime $eE/m\gg m$ for the Schwinger effect for a pure electric field in flat spacetime, i.e., the electric potential energy
across one Compton wavelength of the charge is much larger than the mass of charge. The upper bound for the induced current in the magnetic field and electric field is given by $eB\Omega^{-2}H^{2}/m$ modulo a constant of order one, while in the pure electric field, the induced current has an upper bound given by $eH^{4}/m$, independently of the electric field.
\par

Finally, in the limit of a magnetic field stronger than the mass of charges, the electric field and scalar curvature of the $\ds$, the induced current is proportional to the pseudo-scalar of the Maxwell theory, [see the result in Eq.~(\ref{strong})] which corresponds to the chiral magnetic effect for
spin-$1/2$ fermions \cite{Fukushima:2008xe}. The chiral magnetic effect for fermions in the $\ds$, which is likely to hold for spinor QED considering the analogy with scalar QED, would be physically interesting but is beyond the scope of this paper and will be addressed in a future study.
\par

Going further, an extension of the setups already known to investigate the Schwinger effect in dS would be to consider anisotropic inflationary spacetime where a constant electric field could be naturally sustained.
Links to axion  inflation and possibly a mechanism of baryogenesis with the help of the Schwinger effect could also be exhibited.
%%%%%%%%%%%%%%%%%%%%%%%%%%%%%%%%%%%%%%%%%%%%%%%%%%%%%%%%%%%%%%%%%%%%%%%%%%%%%%%%%%%%%%%%%%%%%%%%%%%%%%%%%%%%%%%%%%%%%%%%%%%%%%%%%%%%%%%%%%%%%%%%%%%%%%%%%%%%%%%
%%%%%%%%%%%%%%%%%%%%%%%%%%%%%%%%%%%%%%%%%%%%%%%%%%%%%%%%%%%%%%%%%%%%%%%%%%%%%%%%%%%%%%%%%%%%%%%%%%%%%%%%%%%%%%%%%%%%%%%%%%%%%%%%%%%%%%%%%%%%%%%%%%%%%%%%%%%%%%%
\acknowledgments
S.~P.~K. would like to thank Remo Ruffini at ICRANet, where this work was initiated and also W-Y.~Pauchy Hwang for the warm hospitality at National
Taiwan University. The work of S.~P.~K. was supported by IBS (Institute for Basic Science) under IBS-R012-D1 and also by the Basic Science Research
Program through the National Research Foundation of Korea (NRF) funded by the Ministry of Education (NRF-2015R1D1A1A01060626).
E.~B. and S.-S.~X. would like to thank S.~P.~K. for the warm hospitality at Kunsan National University.
E.~B. is supported by the University of Kashan.
%%%%%%%%%%%%%%%%%%%%%%%%%%%%%%%%%%%%%%%%%%%%%%%%%%%%%%%%%%%%%%%%%%%%%%%%%%%%%%%%%%%%%%%%%%%%%%%%%%%%%%%%%%%%%%%%%%%%%%%%%%%%%%%%%%%%%%%%%%%%%%%%%%%%%%%%%%%%%%%
%%%%%%%%%%%%%%%%%%%%%%%%%%%%%%%%%%%%%%%%%%%%%%%%%%%%%%%%%%%%%%%%%%%%%%%%%%%%%%%%%%%%%%%%%%%%%%%%%%%%%%%%%%%%%%%%%%%%%%%%%%%%%%%%%%%%%%%%%%%%%%%%%%%%%%%%%%%%%%%
\appendix
\section{\label{app:zeta}Riemann and Hurwitz zeta functions}
In this appendix some useful properties of the Riemann and Hurwitz zeta functions are reviewed; for more properties see, e.g.,~\cite{Book:Nist}.
\par

The Riemann zeta function is a function of the complex variable $s$ that analytically continue the Dirichlet series
\begin{equation}
\zeta\bigl(s\bigr)=\sum_{n=1}^{\infty}\frac{1}{n^{s}}, \label{def1:zeta}
\end{equation}
for when $\Re(s)>1$. Another representation of it is
\begin{equation}
\zeta\bigl(s\bigr)=\frac{1}{1-2^{-s}}\sum_{n=0}^{\infty}\frac{1}{\bigl(2n+1\bigr)^{s}}. \label{def2:zeta}
\end{equation}
Via the analytic continuation of~(\ref{def1:zeta}), it is possible to assign a finite result to the divergent series
\begin{equation}
\sum_{n=1}^{\infty} 1=\zeta\bigl(0\bigr)=-\frac{1}{2}. \label{special}
\end{equation}
Similarly, the Hurwitz zeta function is defined by the series expansion
\begin{align}
\zeta\bigl(s,a\bigr) & =\sum_{n=0}^{\infty}\frac{1}{\bigl(n+a\bigr)^{s}}, & \Re(s)>1,~a\neq 0,-1,-2,\ldots. \label{def:Hurwitz}
\end{align}
The Riemann zeta function is nothing but a special case
\begin{equation}
\zeta\bigl(s,1\bigr)=\zeta\bigl(s\bigr). \label{case}
\end{equation}
The following special values of the Hurwitz zeta function are relevant here
\begin{align}
\zeta\bigl(s,\frac{1}{2}\bigr)& =\bigl(2^{s}-1\bigr)\zeta\bigl(s\bigr), && s\neq 1. \label{special:1} \\
\zeta\bigl(0,a\bigr) &= \frac{1}{2}-a. \label{special:2}
\end{align}
The Hurwitz zeta function satisfies the symmetry of second derivatives and its derivative in the second argument is a shift
\begin{align}
\frac{\rnd^{2}}{\rnd s\rnd a}\zeta\bigl(s,a\bigr)&= \frac{\rnd^{2}}{\rnd a\rnd s}\zeta\bigl(s,a\bigr), \label{derivative:1} \\
\frac{\rnd}{\rnd a}\zeta\bigl(s,a\bigr)&=-s\zeta\bigl(s+1,a\bigr), && s\neq 0,1,2,\ldots;~\Re(a)>0. \label{derivative:2}
\end{align}
One of its limiting behavior reads
\begin{equation}
\lim_{s\rightarrow 1}\Bigl[\zeta\bigl(s,a\bigr)-\frac{1}{s-1}\Bigr]=-\psi\bigl(a\bigr), \label{limit}
\end{equation}
where $\psi(a)$ is the digamma function which has the special value
\begin{equation}
\psi\Bigl(\frac{1}{2}\Bigr)=-\gamma_{E}-\log(4). \label{digamma}
\end{equation}
With Eqs.~(\ref{derivative:1})-(\ref{digamma}), one can verify the useful mathematical formula
\begin{equation}
\frac{\rnd^{2}}{\rnd a\rnd s}\zeta\Bigl(s=0,a=\frac{1}{2}\Bigr)=-\gamma_{\mathrm{Euler}}-\log(4). \label{formula}
\end{equation}
%%%%%%%%%%%%%%%%%%%%%%%%%%%%%%%%%%%%%%%%%%%%%%%%%%%%%%%%%%%%%%%%%%%%%%%%%%%%%%%%%%%%%%%%%%%%%%%%%%%%%%%%%%%%%%%%%%%%%%%%%%%%%%%%%%%%%%%%%%%%%%%%%%%%%%%%%%%%%%%
%%%%%%%%%%%%%%%%%%%%%%%%%%%%%%%%%%%%%%%%%%%%%%%%%%%%%%%%%%%%%%%%%%%%%%%%%%%%%%%%%%%%%%%%%%%%%%%%%%%%%%%%%%%%%%%%%%%%%%%%%%%%%%%%%%%%%%%%%%%%%%%%%%%%%%%%%%%%%%%
\section{\label{app:reg}Adiabatic regularization of the current}
In order to compute the one-dimensional momentum integral in the right hand side of Eq.~(\ref{jreg}), we adopt the integration procedure that have been introduced in \cite{Frob:2014zka,Kobayashi:2014zza}. The Mellin-Barnes representation of the Whittaker function is given by \cite{Book:Nist}
\begin{align}
W_{\kappa,\gamma}(z) &= e^{\frac{-z}{2}} \int_{-i\infty}^{+i\infty}\frac{ds}{2\pi i}
\frac{\Gamma(\frac{1}{2}+\gamma+s)\Gamma(\frac{1}{2}-\gamma+s)\Gamma(-\kappa-s)}{\Gamma(\frac{1}{2}+\gamma-\kappa)\Gamma(\frac{1}{2}-\gamma-\kappa)}
z^{-s}, & \bigl|ph(z)\bigr| & <\frac{3\pi}{2}, \nn \\
\frac{1}{2}\pm\gamma-\kappa & \neq 0,-1,-2,\ldots,\label{MB}
\end{align}
where the contour of integration separates the poles of $\Gamma(\frac{1}{2}+\gamma+s)\Gamma(\frac{1}{2}-\gamma+s)$ from those of $\Gamma(-\kappa-s)$.
Substituting the integral representation~(\ref{MB}) into Eq.~(\ref{jreg}) and choosing the contour of integration similar to \cite{Kobayashi:2014zza},
leads to the final result
\begin{equation}
J=\frac{eH^{3}\ell}{4\pi^{2}}\biggl(\frac{\gamma\sinh\big(2\pi\lambda\big)}{\sin\big(2\pi\gamma\big)}+\lambda\biggr). \label{Jnon}
\end{equation}
\par
In order to regularize the current~(\ref{Jnon}) we apply the adiabatic subtraction method.
Starting from Eq.~(\ref{tpart}), for positive frequency modes it can be rewritten as
\begin{equation}
\frac{d^{2}f_{\A}(\tau)}{d\tau^{2}}+\omega^{2}(\tau)f_{\A}(\tau)=0, \label{mode}
\end{equation}
where the $\omega$ reads
\begin{equation}
\omega(\tau)=+\biggl(k^{2}-\frac{2\lambda k_{z}}{\tau}+\frac{\lambda^{2}+\mu^{2}}{\tau^{2}}-\frac{2}{\tau^{2}}\biggr)^{\frac{1}{2}}. \label{omega}
\end{equation}
A Wentzel-Kramers-Brillouin (WKB) type solution for the mode equation~(\ref{mode}) is
\begin{equation}
f_{\A}(\tau)=\bigl(2W(\tau)\bigr)^{\frac{-1}{2}} \exp\Bigl(-i\int^{\tau} W(\tau')d\tau'\Bigr),  \label{wkb}
\end{equation}
provided that the function $W(\tau)$ satisfies the equation
\begin{equation}
W^{2}(\tau)=\omega^{2}(\tau)+\frac{3\dot{W}^{2}}{4W^{2}}-\frac{\ddot{W}}{2W},  \label{W}
\end{equation}
where the dot indicates a derivative with respect to the conformal time $\tau$.
For the zeroth order adiabatic expansion of $W(\tau)$, the derivative terms on the right hand side of Eq.~(\ref{W}) are neglected, and we then obtain
\begin{equation}
W^{(0)}(\tau)=\omega_{0}^{2}(\tau).  \label{W0}
\end{equation}
Since the last term in Eq.~(\ref{omega}) is of second adiabatic order
\begin{equation}
\frac{2}{\tau^{2}}=2\frac{\dot{\Omega}^{2}}{\Omega^{2}},  \label{second}
\end{equation}
we then have
\begin{equation}
\omega_{0}(\tau)=+\biggl(k^{2}-\frac{2\lambda k_{z}}{\tau}+\frac{\lambda^{2}+\mu^{2}}{\tau^{2}}\biggr)^{\frac{1}{2}}. \label{omega0}
\end{equation}
With Eqs.~(\ref{tilde}),~(\ref{ansatz}),~(\ref{wkb}),~(\ref{W0}), and~(\ref{omega0}) the zeroth adiabatic order for the positive and negative frequency
mode functions are
\begin{eqnarray}
U_{\A}\bigl(x;\ky,n\bigr) &=& \Omega^{-1}(\tau) e^{+i\x\cdot\ky} h(y_{+}) \bigl(2\omega_{0}(\tau)\bigr)^{\frac{-1}{2}} \exp\Bigl(-i\int^{\tau}\omega_{0}(\tau')d\tau'\Bigr),  \label{ua} \\
V_{\A}\bigl(x;-\ky,n\bigr) &=& \Omega^{-1}(\tau) e^{+i\x\cdot\ky} h(y_{+}) \bigl(2\omega_{0}(\tau)\bigr)^{\frac{-1}{2}} \exp\Bigl(+i\int^{\tau}\omega_{0}(\tau')d\tau'\Bigr). \label{va}
\end{eqnarray}
This adiabatic complete set of orthonormal mode functions can be used to construct the Fock space.
Then, the zeroth adiabatic order expansion of the vacuum expectation value of the current operator is given by
\begin{equation}
j_{\A}=e\Omega^{-2}(\tau)\sum_{n=0}^{\infty}\int \frac{d^{2}\dky}{(2\pi)^{2}}\Bigl(k_{z}-\frac{\lambda}{\tau}\Bigr)
\Bigl(|U_{\A}|^{2}+|V_{\A}|^{2}\Bigr). \label{jadi}
\end{equation}
After some algebra and using~(\ref{zeta}), it can be shown that
\begin{equation}
J_{\A}=\Omega(\tau)j_{A}=\frac{eH^{3}\ell\lambda}{4\pi^{2}}. \label{ja}
\end{equation}
The adiabatic regularization scheme consists in subtracting the counterterm~(\ref{ja}) from the original expression~(\ref{Jnon}),
\begin{equation}
J_{\reg}=J-J_{\A}. \label{scheme}
\end{equation}
%%%%%%%%%%%%%%%%%%%%%%%%%%%%%%%%%%%%%%%%%%%%%%%%%%%%%%%%%%%%%%%%%%%%%%%%%%%%%%%%%%%%%%%%%%%%%%%%%%%%%%%%%%%%%%%%%%%%%%%%%%%%%%%%%%%%%%%%%%%%%%%%%%%%%%%%%%%%%%%
%%%%%%%%%%%%%%%%%%%%%%%%%%%%%%%%%%%%%%%%%%%%%%%%%%%%%%%%%%%%%%%%%%%%%%%%%%%%%%%%%%%%%%%%%%%%%%%%%%%%%%%%%%%%%%%%%%%%%%%%%%%%%%%%%%%%%%%%%%%%%%%%%%%%%%%%%%%%%%%

%%%%%%%%%%%%%%%%%%%%%%%%%%%%%%%%%%%%%%%%%%%%%%%%%%%%%%%%%%%%%%%%%%%%%%%%%%%%%%%%%%%%%%%%%%%%%%%%%%%%%%%%%%%%%%%%%%%%%%%%%%%%%%%%%%%%%%%%%%%%%%%%%%%%%%%%%%%%%%%
%%%%%%%%%%%%%%%%%%%%%%%%%%%%%%%%%%%%%%%%%%%%%%%%%%%%%%%%%%%%%%%%%%%%%%%%%%%%%%%%%%%%%%%%%%%%%%%%%%%%%%%%%%%%%%%%%%%%%%%%%%%%%%%%%%%%%%%%%%%%%%%%%%%%%%%%%%%%%%%
\end{document}